
    \def\G{\Gamma}
\def\d{\delta}  \def\ee{\epsilon}
   
 \def\l{\lambda}

 \def\f{\phi} \def\F{\Phi}  
\def\Ps{\Psi}   
\def\pa{\partial}  

%
\newbox\leftpage \newdimen\fullhsize \newdimen\hstitle \newdimen\hsbody
\tolerance=1000\hfuzz=2pt
\def\bigans{b }
\def\answ{b }
%
\ifx\answ\bigans\message{(This will come out unreduced.}
\magnification=1000\baselineskip=16pt plus 2pt minus 1pt
\hsbody=\hsize \hstitle=\hsize 
\else\def\apans{l }\message{ lyman or hepl (l/h) (lowercase]) ? }
\read-1 to \apansw\message{(This will be reduced.}
\let\lr=L
\magnification=1200\baselineskip=16pt plus 2pt minus 1pt
\voffset=-.31truein\vsize=7truein
\hstitle=8truein\hsbody=4.75truein\fullhsize=10truein\hsize=\hsbody
\ifx\apansw\apans\special{ps: landscape}\hoffset=-.59truein
  \else\hoffset=.05truein\fi
\output={\ifnum\pageno=0 
  \shipout\vbox{\hbox to \fullhsize{\hfill\pagebody\hfill}}\advancepageno
  \else
  \almostshipout{\leftline{\vbox{\pagebody\makefootline}}}\advancepageno
  \fi}
\def\almostshipout#1{\if L\lr \count1=1
      \global\setbox\leftpage=#1 \global\let\lr=R
  \else \count1=2
    \shipout\vbox{\ifx\apansw\apans\special{ps: landscape}\fi 
      \hbox to\fullhsize{\box\leftpage\hfil#1}}  \global\let\lr=L\fi}
\fi
%
\catcode`\@=11 
\newcount\yearltd\yearltd=\year\advance\yearltd by -1900

%
%
\def\draftmode{\def\draftdate{{\rm preliminary draft:
\number\month/\number\day/\number\yearltd\ \ \hourmin}}%
\headline={\hfil\draftdate}\writelabels\baselineskip=20pt plus 2pt minus 2pt
{\count255=\time\divide\count255 by 60 \xdef\hourmin{\number\count255}
        \multiply\count255 by-60\advance\count255 by\time
   \xdef\hourmin{\hourmin:\ifnum\count255<10 0\fi\the\count255}}}

\def\nolabels{\def\eqnlabel##1{}\def\eqlabel##1{}\def\reflabel##1{}}
\def\writelabels{\def\eqnlabel##1{%
{\escapechar=` \hfill\rlap{\hskip.09in\string##1}}}%
\def\eqlabel##1{{\escapechar=` \rlap{\hskip.09in\string##1}}}%
\def\reflabel##1{\noexpand\llap{\string\string\string##1\hskip.31in}}}
\nolabels
%
\global\newcount\secno \global\secno=0
\global\newcount\meqno \global\meqno=1
\def\newsec#1{\global\advance\secno by1\message{(\the\secno. #1)}
\xdef\secsym{\the\secno.}\global\meqno=1
\bigbreak\bigskip
\noindent{\bf\the\secno. #1}\par\nobreak\medskip\nobreak}
\xdef\secsym{}
\def\appendix#1#2{\global\meqno=1\xdef\secsym{\hbox{#1.}}\bigbreak\bigskip
\noindent{\bf Appendix #1. #2}\par\nobreak\medskip\nobreak}
%
%
\def\eqnn#1{\xdef #1{(\secsym\the\meqno)}%
\global\advance\meqno by1\eqnlabel#1}
\def\eqna#1{\xdef #1##1{\hbox{$(\secsym\the\meqno##1)$}}%
\global\advance\meqno by1\eqnlabel{#1$\{\}$}}
\def\eqn#1#2{\xdef #1{(\secsym\the\meqno)}\global\advance\meqno by1%
$$#2\eqno#1\eqlabel#1$$}
%
\newskip\footskip\footskip14pt plus 1pt minus 1pt 
\def\f@@t{\baselineskip\footskip\bgroup\aftergroup\@foot\let\next}
\setbox\strutbox=\hbox{\vrule height9.5pt depth4.5pt width0pt}
\global\newcount\ftno \global\ftno=0
\def\foot{\global\advance\ftno by1\footnote{$^{\the\ftno}$}}
%
%
\global\newcount\refno \global\refno=1
\newwrite\rfile
\def\ref{\nref}
\def\nref#1{\xdef#1{[\the\refno]}\ifnum\refno=1\immediate
\openout\rfile=refs.tmp\fi\global\advance\refno by1\chardef\wfile=\rfile
\immediate\write\rfile{\noexpand\item{#1\ }\reflabel{#1}\pctsign}\findarg}
\def\findarg#1#{\begingroup\obeylines\newlinechar=`\^^M\pass@rg}
{\obeylines\gdef\pass@rg#1{\writ@line\relax #1^^M\hbox{}^^M}%
\gdef\writ@line#1^^M{\expandafter\toks0\expandafter{\striprel@x #1}%
\edef\next{\the\toks0}\ifx\next\em@rk\let\next=\endgroup\else\ifx\next\empty%
\else\immediate\write\wfile{\the\toks0}\fi\let\next=\writ@line\fi\next\relax}}
\def\striprel@x#1{} \def\em@rk{\hbox{}} {\catcode`\%=12\xdef\pctsign{
\def\semi{;\hfil\break}
\def\addref#1{\immediate\write\rfile{\noexpand\item{}#1}} 
\def\listrefs{\immediate\closeout\rfile
\baselineskip=14pt\centerline{{\bf References}}\bigskip{\frenchspacing%
\escapechar=` \input refs.tmp\vfill\eject}\nonfrenchspacing}
\def\startrefs#1{\immediate\openout\rfile=refs.tmp\refno=#1}
\def\figures{\centerline{{\bf Figure Captions}}\medskip\parindent=40pt}
\def\fig#1#2{\medskip\item{Fig.~#1:  }#2}
\catcode`\@=12 
%
%
\def\noblackbox{\overfullrule=0pt}
\hyphenation{anom-aly anom-alies coun-ter-term coun-ter-terms}
\def\inv{^{\raise.15ex\hbox{${\scriptscriptstyle -}$}\kern-.05em 1}}
\def\dup{^{\vphantom{1}}}
\def\Dsl{\,\raise.15ex\hbox{/}\mkern-13.5mu D} 
\def\dsl{\raise.15ex\hbox{/}\kern-.57em\partial}
\def\del{\partial}
\def\Psl{\dsl}
\def\tr{{\rm tr}} \def\Tr{{\rm Tr}}
\def\lspace{\ifx\answ\bigans{}\else\qquad\fi}
\def\lbspace{\ifx\answ\bigans{}\else\hskip-.2in\fi} 
\def\boxeqn#1{\vcenter{\vbox{\hrule\hbox{\vrule\kern3pt\vbox{\kern3pt
        \hbox{${\displaystyle #1}$}\kern3pt}\kern3pt\vrule}\hrule}}}
\def\mbox#1#2{\vcenter{\hrule \hbox{\vrule height#2in
                \kern#1in \vrule} \hrule}}  
%
\def\CAG{{\cal A/\cal G}}   
\def\CA{{\cal A}} \def\CC{{\cal C}} \def\CF{{\cal F}} \def\CG{{\cal G}}
\def\CL{{\cal L}} \def\CH{{\cal H}} \def\CI{{\cal I}} \def\CU{{\cal U}}
\def\CB{{\cal B}} \def\CR{{\cal R}} \def\CD{{\cal D}} \def\CT{{\cal T}}
\def\e#1{{\rm e}^{^{\textstyle#1}}}
\def\grad#1{\,\nabla\]_{{#1}}\,}
\def\gradgrad#1#2{\,\nabla\]_{{#1}}\nabla\]_{{#2}}\,}
\def\psibar{\overline\psi}
\def\om#1#2{\omega^{#1}{}_{#2}}
\def\vev#1{\langle #1 \rangle}
\def\lform{\hbox{$\sqcup$}\llap{\hbox{$\sqcap$}}}
\def\darr#1{\raise1.5ex\hbox{$\leftrightarrow$}\mkern-16.5mu #1}
\def\lie{\hbox{\it\$}} 
\def\ha{{1\over2}}
\def\half{{\textstyle{1\over2}}} 
\def\roughly#1{\raise.3ex\hbox{$#1$\kern-.75em\lower1ex\hbox{$\sim$}}}

\ref\gj{J.  Goldstone and R. Jackiw, {\it Phys. Lett.} {\bf
74B}(1978)81.
This approach was also pursued in A. Izergin, M. Semenov, and
L.D. Faddeev, {\it Teor. Mat. Phys.} {\bf 38}(1979)1.}
\ref\ken{K.  Johnson, in {\it QCD -- 20 Years Later}, Aachen, June
1992.}
\ref\luscher{
M. L\"uscher and G. M\"unster, {\it Nucl. Phys.} {\bf B232} (1984) 445.}
\ref\fhjl{D.Z.  Freedman, P.E.  Haagensen, K. Johnson and J.I.  Latorre,
in preparation.}

\pageno=0
\footline={\null \hfill}
\font\lbf=cmbx10 at 12pt
\rightline{\hss UB-ECM-PF 93/16}
\vfill

\centerline{{\lbf On the Exact Implementation of Gauss' Law
in Yang-Mills Theory}\footnote{$^{\dag}$}{Work done in
collaboration with D.Z. Freedman, K. Johnson, and J.I. Latorre.}}

\vskip 6pc

\centerline{Peter E. Haagensen\footnote{*}{E-mail:
{\sl HAGENSEN@EBUBECM1.BITNET.}}}

\def\cl{\centerline}
\cl{\it Departament d'Estructura i Constituents de la
Mat\`eria}
\cl{\it Facultat de F\'\i sica, Universitat de Barcelona}
\cl{\it Diagonal 647, 08028 Barcelona, Spain}

\vskip 6pc

\cl{Lecture given at the {\it XIII Particles and Nuclei}}
\cl{{\it International Conference}, Perugia, Italy, 28 June -- 2 July
1993}

\vskip 5pc

\centerline{\bf Abstract}\vskip 1pc

It is possible to find different sets of local coordinates in
the field space of Yang-Mills theories which implement Gauss' law
manifestly for physical states.  The singular points of the
transformations to these gauge-invariant coordinates induce
energy barriers in the Hamiltonian much like angular momentum in quantum
mechanics, and these formulations suggest a vacuum state functional
qualitatively different from the perturbative one.
\vfill\eject
\footline={\hss \folio \hss}

{\bf 1. Introduction}\vskip .1cm

The implementation of Gauss' law in Hamiltonian QCD in the gauge
$A_0^a=0$ involves a basic difficulty stemming from the non-covariant
transformation law of the vector potential. Evidently, this problem can
be studied in a setting smaller than full QCD, namely, pure nonabelian
gauge theory. Our purpose here is to investigate this constraint for the
case of $SU(2)$ Yang-Mills theory.

One way to solve the problem might be to use a different set of
variables which do transform covariantly under gauge transformations.
For instance, also in $SU(2)$ Yang-Mills theory, Goldstone and Jackiw
proposed some time ago using the electric field as the basic variable
\gj . One might also consider the problem in magnetic variables, and
recently one of us has proposed such an approach \ken .

Here we consider new variables of both magnetic and electric types.
What these new variables help us achieve is a general statement on the
behavior of physical wavefunctionals for slowly-varying fields.
Firstly, we find that the transformation from the original variables
$A_i^a$ to gauge-invariant variables has singular points, and the
resulting Hamiltonian in the new variables will always contain singular
terms at these points.  This is in close analogy to central force
problems in quantum mechanics, where the Hamiltonian has an angular
momentum term which is singular at the origin.  Thus there is an energy
barrier which forces wavefunctionals away from singular points of the
transformation to gauge-invariant variables.  This is in clear
distinction to $\l\f^4$- theory, for instance, or the abelian theory, or
still, the {\it perturbative} nonabelian theory.  Secondly, we find that
these barriers always appear in those terms in the Hamiltonian which
have spatial derivatives, and thus one immediate consequence is that
calculations built around constant-field configurations, e.g. typical
variational methods, will be insensitive to these barriers. \vskip .1cm

{\bf 2. The Magnetic Formulation}\vskip .1cm

In Hamiltonian $SU(2)$ Yang-Mills theory, and in the gauge $A_0^a(x)=0$,
the generator of residual space-dependent gauge transformations is:
\eqn\gen{ G^a(x)=D_kE_k^a(x)=\pa_kE_k^a(x)+\ee^{abc}A_k^b(x)E_k^c(x)\,
.}
$A_i^a(x)$ is the vector potential and $E_i^a(x)$ the electric
field.  In the usual realization of equal-time commutators, $E_i^a
\rightarrow -i\d /\d A_i^a$, the Gauss law constraint
has an inhomogeneous term.  If we consider now the magnetic
field $B_i^a(x)$ as the basic variable\footnote{$^1$}{More rigorously,
this should be calculated as a canonical transformation from $A$- to
$B$-variables.  This is done in \ken .}, it is straightforward to verify
that Gauss' law becomes homogeneous:
\eqn\gtran{ D_i E_i^a (x) = -i D_i {\d\over \d A_i^a (x)}=-i\ee_{ijk}
D_iD_j {\d\over \d B_k^a (x)} =-i\ee^{abc}B_i^b(x){\d\over \d B_i^a (x)}
\, ,}
that is, simply a color rotation when acting on states $\Ps [B]$.
The simplest local quantities which respect
this constraint, and represent the correct number of degrees of
freedom at each space point are:
\eqn\phivar{\f_{ij}=B_i^aB_j^a\longrightarrow G^a\f_{ij}=0\, .}
For our purposes it is
actually convenient to choose another set
of variables, $g_{ij}$, defined by:
\eqn\defg{g_{ij}^{-1}={B_i^aB_j^a\over \det B}\, .}
Wavefunctionals $\Ps [g_{ij}]$ manifestly satisfy Gauss' law, and
the problem then is to consider the measure and Hamiltonian
in these variables.  For lack of time, we will not study the measure in
any detail here, but consider just the Hamiltonian.  The
magnetic energy density term is straightforward
to calculate in the $g$ variables, and we will also not consider it
here.  As for the electric energy density, its expectation on a state
$\Ps [g]$, at a point $x$, will be:
\eqn\esqrd{\eqalign{ <\Ps\vert & E^2\vert\Ps >= -{\d\Ps\over\d A_i^a}
{\d\Ps\over\d A_i^a}=\cr & -\ee_{ijk}\ee_{ij'k'}{g_{uu'}\over\det
g}\left[ \pa_j(2g_{sk}\d_{ut}- g_{st}\d_{uk}){\d\Ps\over\d g_{st}}
-\G_{vju}(2g_{sk}\d_{vt}- g_{st}\d_{vk}){\d\Ps\over\d g_{st}}
\right]\big[{\rm same, with primes} \big]\, ,}}
where $B_i^a\G_{ijk}=D_jB_k^a$.  We do not know the explicit form of
$\G_{ijk}(g)$ for generic potentials, however we do know it for constant
fields, which is the appropriate starting point for the long-wavelength
approximation we want to consider here.  The above expectation, at a
point where $\pa_ig_{jk}=0$, is given by:
\eqn\constg{\eqalign{ <\Ps\vert E^2\vert\Ps >\vert_{\pa g=0}=& -g_{ss'}
{\d\Ps\over\d g_{st}}{\d\Ps\over\d g_{s't}}+ {1\over (\det g)^{1/2}}
{\d\Ps\over\d g_{st}}\left( {\pa\over\pa x_j} {\d\Ps\over\d g_{s't'}}
\right) X(g)_{jsts't'}\cr & +{1\over\det g}\left( {\pa\over\pa x_k}
{\d\Ps\over\d g_{st}}\right)\left( {\pa\over\pa x_{k'}} {\d\Ps\over\d
g_{s't'}} \right) Y(g)_{kk'sts't'} \, .}}
$X$ and $Y$ are lengthy
functions of $g$; all we need to know about them is
that {\it they do not vanish as} $\det g \rightarrow 0$. We note also
that $\pa /\pa x_i\,(\d\Ps /\d g_{jk})\neq 0$
if $\pa_ig_{jk}=0$ only at a point or in a finite region of space.

The above expression already embodies our main claim: the transformation
from $A_i^a$ to $g_{ij}$ is singular when $\det g=0$; there,
in a slowly-varying field approximation, the electric
energy density diverges, and thus there is a barrier
forcing wavefunctionals away from regions of $\det g=0$.  Furthermore,
these barriers appear in terms in the Hamiltonian containing space
derivatives. In
a pure constant-field approximation, such as used in typical variational
methods, only the first term in \constg\ survives\footnote{$^2$}{This
coincides with the Hamiltonian used in \luscher , for instance.}, and
this barrier is not seen at all.
\vskip .1cm

{\bf 3. The Electric Formulation}\vskip .1cm

Here we will only sketch the main steps of our derivation.  We start by
using the realization of canonical equal-time commutators in which
$E_i^a$ rather than $A_i^a$ is diagonal.  The Gauss law operator reads:
\eqn\glelec{ G^a(x)=\pa_kE_k^a(x)-i\ee^{abc}E_k^b(x){\d\over\d E_k^c(x)}
\, .}
It is again inhomogeneous.  Because $E_i^a$ is a $3\times 3$ matrix, it
can always be decomposed as:
\eqn\ets{E_i^a={T^a}_jS_{ji}\, ,}
where $T$ is an orthogonal matrix, and $S$ symmetric.  One can then show
that gauge-invariant wavefunctionals take the form:
\eqn\psie{ \F [E]=e^{-{i\over 2}\int\ee^{abc}E_i^a(\pa_i{T^b}_j)
{T^c}_j}\,\,\F [S] \, .}
By performing a unitary transformation the
above phase can be eliminated, and physical states will be a
function only of $S_{ij}$. It is
possible to verify that Gauss' law becomes homogeneous under this
transformation:
\eqn\glawe{ G^a(x)=-i\ee^{abc}E_i^b(x){\d\over\d E_i^c(x)}=-i \ee^{abc}
T_i^b(x) {\d\over\d T_i^c(x)}\, .}
Furthermore, a lengthy calculation shows that the $A$ field
becomes:\footnote{$^3$}{The field one actually finds is a gauge
transform of this: $\hat{A}_i^a={T^a}_jA_{ji}-1/2\,\ee^{abc}{T^b}_j
\pa_i{T^c}_j$. Since we are only interested in gauge-invariant
quantities, the above simpler form suffices.}
\eqn\abar{ A_{ji}=i[\d_{im}\d_{jn}-\ee_{jiq} V_{qp}\ee_{pmr}
S_{rn}]{\d\over\d S_{mn}}-\ee_{jiq}V_{ql}\pa_mS_{lm}\, ,}
apart from a color rotation in $T$ which we have dropped.  The matrix
$V$ above is:
\eqn\vee{V_{ij}^{-1}=\tr S\,\d_{ij}-S_{ij}\, .}
The quantity $\det V^{-1}$ also appears in the measure. $V$ is singular
whenever the sum of any two eigenvalues of $S$ goes to 0. In those
regions the measure vanishes and the transformation is singular.  Apart
from some normal ordering subtleties involved in dropping the color
rotation in $A$, the situation we now have is the following: the
1-space-derivative term in $A$ has a singularity when two eigenvalues of
$S$ add to 0. The 0-derivative term also appears to have this
singularity, however, it can be seen that $J=0$ states (i.e., states
which are rotation invariant and functions of only three variables: $\tr
S$, $\tr S^2$ and $\tr S^3$), are entirely blind to it.  Therefore, we
are again led to the statement that, in going to the gauge-invariant
variables $S_{ij}$, there are singular regions where the magnetic energy
density will become infinite, and the energy barriers in the Hamiltonian
will (for $J=0$ states) appear in space-derivative terms.

Due to limitations in both time and space, this report has been somewhat
condensed.  A more detailed report with further work along these lines
is in preparation\fhjl .\vskip .1cm

\noindent{\bf Acknowledgments}

It is a pleasure to thank my collaborators, D.Z.  Freedman, K.
Johnson and J.I.  Latorre, for all the discussions and insight which led
to this work.
\vskip .1cm

\listrefs \vfil\eject \end